\documentclass[aps,pre,reprint,twocolumn, nofootinbib, superscriptaddress, longbibliography]{revtex4-2}

\usepackage{color,amsthm,amsmath,amsfonts,graphicx,bm}

\usepackage[utf8]{inputenc}
\usepackage[breakable]{tcolorbox}
\usepackage{xcolor}
\usepackage{algorithm}
\usepackage[noend]{algpseudocode}
\usepackage{wrapfig}
\usepackage{appendix}
\usepackage{amsmath}
\usepackage{amsthm}
\usepackage{amsfonts}
\usepackage{graphicx}
\usepackage{subcaption}
\usepackage{amssymb}
\usepackage{dsfont}
\usepackage{mathtools}
\usepackage{enumerate}
\usepackage{braket}
\usepackage{tikz}
\usetikzlibrary{quantikz2}
\usepackage{wrapfig}
\usepackage{tikz-cd}

\newcommand{\dmat}[1]{{| #1 \rangle \langle #1 |}}

\usepackage{cleveref}

\DeclareMathOperator{\Tr}{\textnormal{Tr}}

\begin{document}
\title{Measuring Butterfly Velocity in the XY Model on Emerging Quantum Computers}
\author{Calum McCartney}
\email{ckm38@cam.ac.uk}
\affiliation{Department of Applied Mathematics and Theoretical Physics, University of Cambridge, UK}
\author{Eric Chen}
\email{ecchen@illinois.edu}
\affiliation{Department of Mathematics, University of Illinois Urbana-Champaign, USA}
\author{Subhayan Roy Moulik}
\email{roymoulik@damtp.cam.ac.uk}
\affiliation{Department of Applied Mathematics and Theoretical Physics, University of Cambridge, UK}

\begin{abstract}
        The butterfly velocity is commonly used to understand information transport properties in quantum dynamical systems and is related to growth of operators. Here we utilise a quantum teleportation based protocol and Riemannian Trust-Region method to estimate the butterfly velocity via the operator averaged out-of-time-order correlation function. We particularly study the XY model and analytically find the maximum group velocity. We then report a proof-of-concept demonstration of this method to estimate the butterfly velocity on NISQ-devices. The numerical simulation results obtained here are compared with our analytical calculations and found to be in agreement. The quantum algorithmic methods presented here can be more generally utilised to study information transport properties in more complicated lattice models.
\end{abstract}

\maketitle

\section{Introduction} 
The butterfly velocity is a measure of the speed at which a small perturbation spreads in a quantum system, and gives a bound on the information spreading velocity~\cite{RS16, MSS16}. The butterfly velocity prominently offers a stronger bound than the celebrated Lieb-Robinson bound for systems with local interactions~\cite{LR72}, in the sense that it may be state dependent~\cite{HH19} and asymmetrically growing in different directions of information propagation~\cite{LGDGG18, SKH18, ZK20}.  The butterfly velocity is commonly used to understand transport properties, such as in the study of behaviour of materials~\cite{B16,Q17, HHM17, L17, BMEK17, PCSS17}. It can be characterised by growth of local operators over time~\cite{AMPSS13, K15, HKLMS21}.

For a system of $n$ qubits, described by state $\rho$, evolving under Hamiltonian $H$, the growth of local operators $W$ ($V$), acting on subspaces labelled $j$ ($1$), is diagnosed by measuring growth of the squared commutator, 
\begin{equation}
    C_{j}(t) \equiv \frac{1}{16} \sum_{W,V} \Tr \left(\rho |[V_1,W_j(t)]|^2\right) = 2-2 \text{Re} \braket{\overline{\text{OTOC}}}_{\rho},
\end{equation}
where $\braket{\overline{\text{OTOC}}}_{\rho}$ is the operator-averaged out-of-time-ordered correlation function  \cite{LO69, K15, S18}, and is defined as,  $\braket{\overline{\text{OTOC}}}_{\rho} \equiv \frac{1}{16}\sum_{W,V}\braket{W(t)^{\dagger}V^{\dagger}W(t)V}_{\rho}$. Here $W$,$V$ act on local spin sites and are averaged over the set of usual Pauli operators. $W(t)= e^{-iHt} W e^{iHt}$ is defined as in the Heisenberg interaction picture, and the expectation value is taken over the state $\rho$. The failure of commutativity for initially commuting operators in time then probes the growth of operators and diagnoses the butterfly velocity.

In this work, we present a method based on adaptation of the YKY protocol \cite{YK17, YY19} which gives a prescription to measure $\braket{\overline{\text{OTOC}}}_{\rho}$ by effectively teleporting information between two subspaces of a dynamically evolving state. The teleportation fidelity then estimates and upper bounds $\braket{\overline{\text{OTOC}}}_{\rho}$, and the rate of decay of OTOCs across qubits is then used to diagnose the growth of commutators and measure the butterfly velocity. A prominent advantage of the YKY algorithm to measure OTOCs is its robustness to certain noise effects as analysed in \cite{YY19}, without requiring explicit error mitigation techniques \cite{Qothers24}.

 A crucial subroutine used in the YKY algorithm to dynamically evolve the state is Hamiltonian simulation. This is realised here using a numerical Hamiltonian-to-circuit mapping technique based on manifold optimisation through the so-called Riemannian trust-region (RTR) method \cite{ABG07, B23}. The RTR method offers a significant depth reductions over Lie-Trotter-Strang type splitting methods and can be scaled for larger systems using classical preprocessing \cite{HVMH21,RVP24}. We report here a proof-of-concept demonstration of the YKY-RTR method to measure the operator-averaged OTOC and subsequently the butterfly velocity.

%XY Model
We particularly study the 1d anisotropic XY Model with a transverse magnetic field term as a simple toy model. This spin lattice model is given by the Hamiltonian,
\begin{equation}
\label{H_XY}
    H = J\sum_j \left(\frac{1+r}{2}X_jX_{j+1} +\frac{1-r}{2} Y_jY_{j+1} + hZ_j \right)
\end{equation}
We note, for Hamiltonian \eqref{H_XY}, the butterfly velocity is state independent. Furthermore, this model can be transformed into a fermionic Hamiltonian and exactly diagonalised, which allows for an analytical expression for the butterfly velocity to be computed as the maximum quasiparticle group velocity. We analytically compute this butterfly velocity for the Hamiltonian \eqref{H_XY} in Section \ref{SEC:Analytic} and report the numerical OTOC calculations obtained from simulations of quantum computers in section \ref{Results} and find them to be in agreement with the analytical obtained predictions. 

\newpage 
\section{Analytical Calculations for Butterfly Velocity in the XY model}\label{SEC:Analytic}
To analytically calculate the butterfly velocity for the XY model, we first perform a Jordan-Wigner transform to convert the spin lattice Hamiltonian (1) into a fermionic Hamiltonian. The convention we will use for this transform in this paper is:
\begin{align}
    X_j &= - \prod_{k<j}(\mathbb{I}-2f_k^{\dagger} f_k)(f_j+f_j^{\dagger}) \nonumber \\
    Y_j &= -i \prod_{k<j}(\mathbb{I}-2f_k^{\dagger} f_k)(f_j-f_j^{\dagger}) \\
    Z_j &= \mathbb{I}-2f_j^{\dagger} f_j \nonumber
\end{align}

Here $f_j, f_j^{\dagger}$ are the fermion annihilation and creation operators for the site $j \in \{1, ..., n\}$. They satisfy the anti-commutation rules $\{f_i, f_j^{\dagger} \} = \delta_{i, j} \mathbb{I}$ and $\{f_i, f_j \} = 0$, where $\{A, B\} = AB + BA$. Ignoring a boundary term (which will be irrelevant in the large $n$ limit), our XY model hence becomes:
\begin{align}
    H = J\sum_{j} &r(f_{j+1}f_{j} + f_j^{\dagger}f_{j+1}^{\dagger}) + (f_{j+1}^{\dagger} f_j + f_j^{\dagger} f_{j+1}) \nonumber \\
    &+ h (\mathbb{I}-2f_j^{\dagger} f_j) 
\end{align}

This form of the Hamiltonian has terms of the form $f_jf_{j+1}$ which don't conserve fermion number. Therefore, the next step is to transform into momentum space by Fourier transforming each fermionic position operator $f_j$ into a non-local momentum operator via $f_j  =  \sum_{k=1}^n c_k e^{ikj}/\sqrt{n}$ (the lattice spacing $a$ is taken as 1 under convention with the usual spin Hamiltonian). In the momentum space the Hamiltonian thus becomes:
\begin{align}
     H = -J \sum_k & \enspace 2(h-\cos(k)) c_k^{\dagger}c_k \\
     &+ ir\sin(k) (c_{-k}^{\dagger}c_k^{\dagger} + c_{-k}c_k)-h\mathbb{I} \nonumber
\end{align}

The final step is a Bogoliubov transform of the form $\gamma_k = u_k c_k -iv_k c_{-k}^{\dagger}$ to rotate into a basis which removes terms that don't conserve fermion number. To ensure the conservation of fermionic anti-commutation relations requires $u_k^2+v_k^2=1$, $u_{-k} = u_k$ and $v_{-k} = -v_k$, prompting us to write $u_k = \cos(\theta_k/2)$ and $v_k = \sin(\theta_k/2)$. This gives us the Hamiltonian:
\begin{align}\label{Bogoliubov}
    H  = &-J \sum_k \gamma_k^{\dagger}\gamma_k[2(h-\cos{ka})\cos^2{\frac{\theta_k}{2}}+r\sin{ka}\sin{\theta_k}] \nonumber \\
    &+ \gamma_{-k}\gamma_{-k}^{\dagger}[2(h-\cos{ka})\sin^2{\frac{\theta_k}{2}}-r\sin{ka}\sin{\theta_k}] \nonumber \\ 
    &+ i(\gamma_{-k}\gamma_{k} + \gamma_{-k}^{\dagger}\gamma_k^{\dagger})(r\sin{ka}\cos{\theta_k} -(h-\cos{ka})\sin{\theta_k}) \nonumber \\ 
    &-h\mathbb{I} 
\end{align}

By choosing the Bogoliubov angle $\theta_k$ such that $\tan{\theta_k} = r\sin{ka}/(h-\cos{ka})$ we eliminate the third term in \eqref{Bogoliubov}. Shifting the other terms around (described in more detail in Appendix \ref{SEC:APPENDIX}) we reach the fully diagonalised Hamiltonian:
\begin{equation}
    H = \sum_k \varepsilon(k;J,r,h) \left(\gamma_k^{\dagger} \gamma_k -\frac{1}{2}\right)
\end{equation}

with the energy dispersion relation:
\begin{equation}\label{Energy}
    \varepsilon(k;J,r,h) = -2J\sqrt{(h-\cos{k})^2+r^2\sin^2{k}}
\end{equation}

To find the butterfly velocity, we then first differentiate \eqref{Energy} with respect to the momentum $k$ to acquire the group velocity,

\begin{equation}\label{v_g}
    v_g(k;J,r,h) = -2J \frac{\sin{k}(h-\cos{k})+r^2\sin{k}\cos{k}}{\sqrt{(h-\cos{k})^2+r^2\sin^2{k}}}
\end{equation}

Following work done in \cite{ML24}, the butterfly velocity is expected to match the maximum of \eqref{v_g} over $k$. For certain special cases such as the Isotropic model ($r=0$) or the TFIM model ($r=1$) we can solve this exactly to find an analytic expression for the butterfly velocity. For arbitrary values of the Hamiltonian parameters however, it seems that no such closed form is easily found. We used graphical methods for calculations of the butterfly velocity on given parameters $r$ and $h$, and these values can be found in \Cref{Table:Params}.

\begin{figure}[H]
    \centering
    \includegraphics[height=60mm]{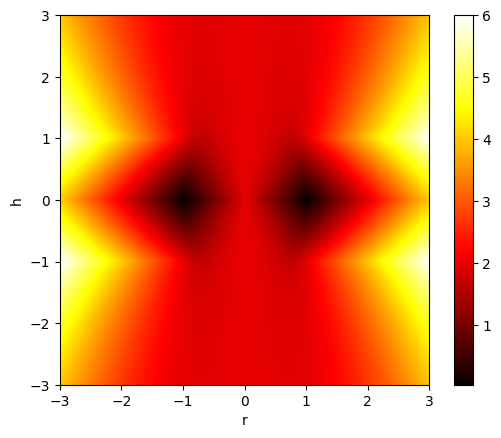}
    \caption{Butterfly velocity, $v_B = \max_k v_g(k; J, r,h)$ for $J=1$, over parameters $r,h$}\label{GroupVelocities} 
\end{figure}

\Cref{GroupVelocities} shows a colour map of butterfly velocities $v_B(r, h)$ (taking $J=1$ to match energy and time units) over a range of the two parameters. It is clear from this plot that $v_B$ has symmetry under parity transforms $r \rightarrow -r$ and $h \rightarrow -h$. Furthermore, the butterfly velocity in the XY-model is state independent and subsequently it suffices to study infinite temperature states containing a uniform mixture of all eigenstates.  

\newpage
\section{Algorithmically estimating  Butterfly Velocity}
\subsection{YKY Algorithm}
\label{SEC:YKY}
For spin-1/2 lattice systems, the butterfly velocity may be alternately estimated via operator averaged out-of-time-order correlation functions. The YKY Algorithm is a bounded error quantum algorithm, originally introduced  and analysed in  \cite{YK17, YY19} that (robustly) measures the operator-averaged OTOC, $   \braket{\widetilde{\text{OTOC}}} \equiv \iint_{Haar} \langle W^{\dagger}(t) V^{\dagger} W(t) V \rangle dVdW$. Here $\int dV$ is the Haar average over all unitary operators on subsystem $V$, and $W(t) = U_t^{\dagger}WU_t$, where $U_t = e^{-iHt}$ is the unitary dynamics. The expectation value is taken over all eigenstates, or equivalently the maximally mixed state on n qubits, $\rho = \frac{I}{2^n}$. Specialising this setup to spin-$1/2$ systems and probing individual lattice sites $1$ and $j \in [n]$, this Haar integral can be replaced by an average over Pauli operators as,
\begin{equation}
    \label{OTOC_Avg}
    \braket{\overline{\text{OTOC}}} = \frac{1}{16} \sum_{\substack{V_1 \in \{ \mathbb{I}, X_1, Y_1, Z_1 \},\\ W_j \in \{ \mathbb{I}, X_j, Y_j, Z_j \}}}  \langle W_j(t) V_1 W_j(t) V_1 \rangle.
\end{equation}

This $\braket{\overline{\text{OTOC}}} = \frac{1}{16}\sum_{V,W}Tr[W_j(t)V_1W_j(t)V_1 \frac{I}{2^n}]$, with local and commuting $W,V$, can be diagrammatically expressed as,
\begin{equation} \frac{1}{16}\sum_{V,W}
    \begin{quantikz}[row sep = 0.12cm, column sep = 0.2cm]
    \makeebit[angle=0]{} & & & & & & & & & \makeebit[angle=0]{} \\
     & \gate{V} & \gate[2, style={draw=blue}]{U} & & \gate[2, style={draw=green}]{U^\dagger} & \gate{V} & \gate[2, style={draw=red}]{U} & & \gate[2, style={draw=yellow}]{U^\dagger} & \\
    \makeebit[angle=0]{} & & & \gate{W} & & & & \gate{W} & & \makeebit[angle=0]{} \\
     & & & & & & & & &
\end{quantikz}
\label{diag_OTOC}
\end{equation}
Using diagrammatic calculus \cite{Penrose06, CK18} and linearity rules for \ \ \ \ $(U\otimes I)\sum_{j=1}^d \frac{\ket{j}\otimes \ket{j}}{\sqrt{d}} = (I\otimes U^T)\sum_{j=1}^{d}\frac{\ket{j}\otimes \ket{j}}{\sqrt{d}}$, 
\begin{gather*}
\begin{quantikz}[align equals at = 1.5]
    \makeebit[angle=0]{} & \gate{U} &  \\
     & &
\end{quantikz}
\, =
\begin{quantikz}[align equals at = 1.5]
    \makeebit[angle=0]{} & &  \\
     & \gate{U^T} &
\end{quantikz},
\end{gather*}
expression \eqref{OTOC_Avg} can be equivalently expressed as,
\begin{equation} 
    \label{OTOC_eq5}
    \frac{1}{16}\sum_{V,W}
    \begin{quantikz}[row sep = 0.12cm, column sep = 0.2cm] 
    \makeebit[angle=0]{} & \gate{V} & \gate[2, style={draw=blue}][0.79cm]{U} & & \gate[2, style={draw=green}][0.79cm]{U^\dagger} & & & \makeebit[angle=0]{}\\
    \makeebit[angle=0]{} & \wireoverride{n}\makeebit[angle=0]{} & & \gate{W} & & & \makeebit[angle=0]{} & \wireoverride{n}\makeebit[angle=0]{} \\
    \makeebit[angle=0]{} & \wireoverride{n} &  \gate[2, style={draw=yellow}][0.79cm]{U^*} & \gate{W^T} & \gate[2, style={draw=red}]{U^T} & & & \wireoverride{n}\makeebit[angle=0]{} \\
    & & & & & & \gate{V^T} &
\end{quantikz}
\end{equation}
\begin{equation*}
%\begin{gather*}
\text{Lastly, using relation, } \sum_{ P \in \{I,X,Y,Z\}} \frac{1}{4}
 P^T \otimes P = 
\begin{quantikz}
    & \makeebit[angle=0, label style = {xshift = 0.28cm}]{} & \setwiretype{n}\makeebit[angle=0, label style = {xshift = 0.28cm}]{} & \setwiretype{q} \\
    & & \setwiretype{n} & \setwiretype{q}
\end{quantikz}
%\end{gather*} 
\end{equation*}
Expression \eqref{OTOC_eq5} simplifies as,
\begin{equation} 
    \begin{quantikz}[row sep = 0.15cm, column sep = 0.2cm] 
    \makeebit[angle=0]{} & & & & & & \makeebit[angle=0]{} \\
    & \gate[2, style={draw=blue}][0.79cm]{U} & & & & \gate[2, style={draw=green}][0.79cm]{U^\dagger} & \\
    \makeebit[angle=0]{} & & \makeebit[angle=0]{} & \setwiretype{n}\makeebit[angle=0]{} & \setwiretype{q} & & \makeebit[angle=0]{}\\
    & \gate[2, style={draw=yellow}][0.79cm]{U^*} & & \wireoverride{n} & \wireoverride{q} & \gate[2, style={draw=red}]{U^T} & \\
    \makeebit[angle=0]{} & & & & & & \makeebit[angle=0]{} \\
    & & & & & & 
\end{quantikz} = 
%P_{\text{EPR}} = 
\bra{\Psi_{\text{f}}}(I \otimes \Pi_{A_jB_j} \otimes I) \ket{\Psi_{\text{f}}}
\label{diag_Pepr}
\end{equation}

\begin{flalign}
\text{where, } \ket{\Psi_{\text{f}}} =   (I_{A_0}\otimes U \otimes U^{*}\otimes I_{B_0})\ket{\Psi}  \\
    \ket{\Psi} = \ket{\phi}_{A_0A_1} \ket{\phi}_{A_2B_2}  \ldots \otimes \ket{\phi}_{A_nB_n} \ket{\phi}_{B_0B_1}, \nonumber \\
    \ket{\phi}_{AB} = \frac{\ket{00}_{AB} + \ket{11}_{AB}}{\sqrt{2}}, \ \  \Pi_{A_jB_j} = \dmat{\phi}_{A_jB_j} \nonumber
        \label{psi_in}
\end{flalign}

$\braket{\overline{\text{OTOC}}}$ (eq \eqref{OTOC_Avg}) can then be measured using the quantum circuit given in Figure \ref{OTOC_circuit1}, by measuring the probability of obtaining outcomes "00", on qubits $A_j,B_j$ as $\braket{\overline{\text{OTOC}}} = \bra{\Psi_{\text{f}}}(I \otimes \Pi_{A_jB_j} \otimes I) \ket{\Psi_{\text{f}}}$, as shown in \eqref{diag_Pepr}. While this process is algorithmically correct in the idealised noiseless computations, more detailed analysis of this process in \cite{YK17,YY19}, suggests an alternate method to estimating OTOCs by instead estimating the probability, $F_{\text{EPR}}$, of  outcomes "00", on qubits $A_1,B_1$, conditioned on having obtained outcomes "00" on qubits $A_jB_j$, as,
\begin{equation}
    F_{\text{EPR}} = \frac{1}{\braket{\overline{\text{OTOC}}}}\bra{\Psi_{\text{f}}} \Pi_{A_jB_j} \Pi_{A_1B_1} \Pi_{A_jB_j} \ket{\Psi_{\text{f}}} = \frac{1}{4 \braket{\overline{\text{OTOC}}}}
    \label{Fepr1}
\end{equation}    

The error bounds on the estimated ${\braket{\overline{\text{OTOC}}}}$ as $\frac{1}{4F_{\text{EPR}}}$ has been previously analysed in \cite{YK17} and shown robust to certain noise effects such as decoherence and small coherent errors, in the sense that $\frac{1}{4F_{\text{EPR}}}\geq \braket{\overline{\text{OTOC}}}$.
The equality, \eqref{Fepr1}, can be verified using again the diagrammatic calculus, starting with, 

\begin{equation}
\ket{\Psi_{\text{f}}} =
    \begin{quantikz}[row sep = 0.15cm, column sep = 0.2cm] 
    \makeebit[angle=0]{} & &  \\
    & \gate[2, style={draw=blue}][0.79cm]{U} & \\
    \makeebit[angle=0]{} & & \\
    & \gate[2, style={draw=yellow}][0.79cm]{U^*} & \\
    \makeebit[angle=0]{} & &  \\
    & & 
\end{quantikz}
\label{Psi_f}
\end{equation}
 
\noindent
Using relation,  $\Pi_{A_jB_j} = \dmat{\phi}_{A_jB_j} =  
\begin{quantikz}
    {\tiny A_j} &  \makeebit[angle=0, label style = {xshift = 0.28cm}]{} & \setwiretype{n}\makeebit[angle=0, label style = {xshift = 0.28cm}]{} & \setwiretype{q} \\
    {\tiny B_j} & & \setwiretype{n} & \setwiretype{q}
\end{quantikz}$, 
measuring qubits $A_j, B_j$ in the Bell basis, and projecting the state $\ket{\Psi_{\text{f}}}$ onto $\Pi_{A_jB_j}$, the (normalised) post-measurement state then becomes,

\begin{equation} 
\frac{\Pi_{A_jB_j}\ket{\Psi_{\text{f}}}}{{\sqrt{\bra{\Psi_{\text{f}}}(\Pi_{A_jB_j}) \ket{\Psi_{\text{f}}}}}} = \frac{1}{\sqrt{\braket{\overline{\text{OTOC}}}}}
    \begin{quantikz}[row sep = 0.15cm, column sep = 0.2cm] 
    \makeebit[angle=0]{} & & & &  \\
    & \gate[2, style={draw=blue}][0.79cm]{U} & & & \\
    \makeebit[angle=0]{} & & \makeebit[angle=0]{} & \setwiretype{n}\makeebit[angle=0]{} & \setwiretype{q} \\
    & \gate[2, style={draw=yellow}][0.79cm]{U^*} & & \wireoverride{n} & \wireoverride{q} \\
    \makeebit[angle=0]{} & & & & \\
    & & & & 
\end{quantikz}
\label{PiPsi_f}
\end{equation}
Finally, measuring qubits $A_1, B_1$ in the Bell basis again, the probability of projecting the state onto $\Pi_{A_1B_1}$ and obtaining outcomes "00" can be given as, 
\begin{align}
    F_{\text{EPR}} = & \frac{1}{4 \braket{\overline{\text{OTOC}}}}
\begin{quantikz}[row sep = 0.15cm, column sep = 0.2cm] 
    \makeebit[angle=0]{} & & & & \makeebit[angle=0, style = dotted]{} & \setwiretype{n} \makeebit[angle=0, style = dotted]{} & \setwiretype{q} & & & & \makeebit[angle=0]{} \\
    & \gate[2, style={draw=blue}][0.79cm]{U} & & & \makeebit[angle=0, style = dotted]{} & \makeebit[angle=0, style = dotted]{} & & & & \gate[2, style={draw=green}][0.79cm]{U^\dagger} & \nonumber \\
    \makeebit[angle=0]{} & & \makeebit[angle=0]{} & \setwiretype{n}\makeebit[angle=0]{} & \setwiretype{q}\makeebit[angle=0, style = dotted]{} & \makeebit[angle=0, style = dotted]{} & \makeebit[angle=0]{} & \setwiretype{n}\makeebit[angle=0]{} & \setwiretype{q} & & \makeebit[angle=0]{}\\
    & \gate[2, style={draw=yellow}][0.79cm]{U^*} & & \wireoverride{n} & \wireoverride{q}\makeebit[angle=0, style = dotted]{} & \makeebit[angle=0, style = dotted]{} & & \wireoverride{n} & \wireoverride{q} & \gate[2, style={draw=red}]{U^T} & \\
    \makeebit[angle=0]{} & & & & \makeebit[angle=0, style = dotted]{} & \makeebit[angle=0, style = dotted]{} & & & & & \makeebit[angle=0]{} \\
    & & & & & \wireoverride{n} & & \wireoverride{q} & & &
\end{quantikz} \nonumber \\
 &  = \frac{1}{4 \braket{\overline{\text{OTOC}}}}
    \begin{quantikz}[row sep = 0.12cm, column sep = 0.2cm] 
    \makeebit[angle=0]{} & & \gate[2, style={draw=blue}][0.79cm]{U} & & & \gate[2, style={draw=green}][0.79cm]{U^\dagger} & & \makeebit[angle=0]{}\\
    \makeebit[angle=0]{} & \wireoverride{n}\makeebit[angle=0]{} & & \makeebit[angle=0]{} & \setwiretype{n}\makeebit[angle=0]{} & \setwiretype{q} & \makeebit[angle=0]{} & \wireoverride{n}\makeebit[angle=0]{} \\
    \makeebit[angle=0]{} & \wireoverride{n} &  \gate[2, style={draw=yellow}][0.79cm]{U^*} & & \wireoverride{n} &  \gate[2, style={draw=red}]{U^T} & & \wireoverride{n}\makeebit[angle=0]{} \\
    & & & & & & &
\end{quantikz} \nonumber \\
 & = \frac{1}{4 \braket{\overline{\text{OTOC}}}}
\end{align}

as in \eqref{Fepr1}. This $F_{\text{EPR}}$ can be equivalently measured using a quantum circuit given in \Cref{OTOC_circuit1}. 

Furthermore, from the analysis in \cite[Proposition 3]{SAZ21}, it follows that to estimate $F_{\text{EPR}}$ up to $\epsilon$ additive precision,  it suffices to only have $O(\epsilon^2)$ samples, as the probability that a random sample deviates significantly from the mean value is exponentially small. 

\begin{figure}[H]
\hspace{-0.5cm}
    \begin{quantikz}[row sep = 0.15cm, column sep = 0.2cm]
    \lstick{$A_0 \ket{0}$} & \gate{H} & \ctrl{1} & & & & & & & & \ctrl{9} & \gate{H} & \meter{}\\
    \lstick{$A_1 \ket{0}$} & & \targ{} & & \gate[4, style={draw=blue}][1.5cm]{U}[1cm]\gateinput{1} & & & & & & & & \\
    \lstick{$A_2 \ket{0}$} & \gate{H} & \ctrl{5} & & \gateinput{2} & & & & & & & & \\[-0.2cm]
    \lstick{\vdots} \setwiretype{n} & \ \vdots\ & & & \gateinput[label style = {yshift = -0.2cm}]{\vdots} & & & & & & & & \\
    \lstick{$A_n \ket{0}$} & \gate{H} & & \ctrl{1} & \gateinput{n} & \ctrl{1} & \gate{H} & \meter[label style = {xshift = 0.7cm, yshift = -0.5cm}]{= 0}\\
    \lstick{$B_n \ket{0}$} & & & \targ{} & \gate[4, style={draw=yellow}][1.5cm]{U^*}[1cm]\gateinput{n} & \targ{} & & \meter[label style = {xshift = 0.7cm, yshift = -0.5cm}]{= 0}\\[-0.25cm]
    \lstick{\vdots} \setwiretype{n} & \ \vdots\ & & & \gateinput[label style = {yshift = -0.2cm}]{\vdots} & & & & & & & & \\
    \lstick{$B_2 \ket{0}$} & & \targ{} & & \gateinput{2} & & & & & & & & \\
    \lstick{$B_1 \ket{0}$} & \gate{H} & \ctrl{1} & & \gateinput{1} & & & & & & & & \\
    \lstick{$B_0 \ket{0}$} & & \targ{} & & & & & & & & \targ{} & & \meter{}
    \end{quantikz}
    \caption{Quantum circuit encoding $U=e^{-iHt}$ to estimate $\braket{\overline{\text{OTOC}}}$ on the doubled Hilbert space of $H$}
    \label{OTOC_circuit1}
\end{figure}
\noindent

\subsection{Manifold Optimisation for Hamiltonian to Circuit Mapping}
\label{SEC:RTR}

In applying the YKY algorithm to the Hamiltonian \eqref{H_XY}, a necessary step is to construct a circuit approximating $U=e^{-iHt}\in U(2^n)$. Here we use Riemannian Trust Region method to construct a circuit approximating $e^{-iHt}$ using $U(4)$ gates. 

Riemannian Trust-Region is an optimisation method similar in application to gradient descent, but with much faster optimisation \cite{ABG07, AMS08, B23}. In general, given a Riemannian manifold $(\mathcal{M},g)$ equipped with a retraction function $R:T\mathcal{M}\rightarrow\mathcal{M}$, where $T\mathcal{M}$ is the tangent bundle of $\mathcal{M}$, and a cost function $f:\mathcal{M}\rightarrow\mathbb{R}$, RTR aims to find local minima of $f$ by iteratively solving a series of local minimisation problems. More precisely, given a current iterate $x_k\in\mathcal{M}$ and trust-region radius $\Delta_k>0$, the method minimises the second-order approximation
\begin{equation}
\hat{m}_{x_k}(\eta)=f(x_k)+\langle (\nabla f)_{x_k},\eta\rangle+\frac{1}{2}(\nabla^2 f)_{x_k}(\eta,\eta)
\end{equation}
of $(R_{x_k})^*f:T_{x_k}\mathcal{M}\rightarrow\mathbb{R}$ over $\eta\in T_{x_k}\mathcal{M}$ in the trust-region $\{|\eta|\leq\Delta_k\}$. Here $R_{x_k}$ denotes $R(x_k,\cdot):T_{x_k}\mathcal{M}\rightarrow\mathcal{M}$, and all inner products and covariant derivatives are computed on $T_{x_k}\mathcal{M}$ with respect to the pullback metric $(R_{x_k})^*g$. If $\eta_k$ is a solution (or an approximate solution) to this trust-region subproblem, then one first computes the quotient
\begin{equation}
    \rho_k=\frac{f(x_k)-f(R_{x_k}(\eta_k))}{\hat{m}_{x_k}(0)-\hat{m}_{x_k}(\eta_k)},
\end{equation}
which describes how well the second-order local minimisation problem associated with $\hat{m}_{x_k}$ approximates the local minimisation problem for $f$. Roughly, if $\rho_k\ll 1$ then this approximation is poor and our trust-region is too large; we keep the same iterate for the next step and decrease the trust-region radius by setting $x_{k+1}=x_k$ and $\Delta_{k+1}=\frac{1}{4}\Delta_k$. Otherwise, the approximation is fair or at least produces a large decrease in $f$, so $\Delta_{k+1}$ can either be kept the same as $\Delta_k$ or increased, and the next iterate is defined as $x_{k+1}=R_{x_k}\eta_k$. Additional details can be found in \cite{ABG07}. 

Under suitable hypotheses on the cost function $f$ and the retraction $R$, the sequences $\{x_k\}_{k=1}^\infty$ produced by the RTR method are known to converge to critical points of the cost function for all choices of initial iterates $x_1$ and trust-region radii, in contrast to the Newton method, which does not converge in some cases. In particular, smoothness of $f$ together with compactness of $\mathcal{M}$ is a sufficient hypothesis to give this global convergence to critical points. Although RTR will not always give convergence to a local minimum of $f$, numerical experiments have suggested that it does so generically. Moreover, the algorithms used to solve the trust-region subproblems at each iteration step converge superlinearly. The restriction to a local minimisation subproblem at each iteration step of RTR also offers a computational advantage compared to usual Newton method.

In our case, we apply RTR in the setting of a specific brick-wall circuit of a preselected depth $m$, where each layer contains a single two-qubit gate applied to alternating pairs of qubits. The figure below illustrates the brick wall circuit $E(G_1,G_2,\ldots,G_m)\in U(2^n)$ generated by $G=(G_1, G_2,\ldots,G_m)\in U(4)^m$, in the case $n=6$ and $m=4$.

\begin{equation*}
\begin{quantikz}[row sep = 13.2]
     & \gate[6,style={draw=blue}]{U} & \\
     & & \\
     & & \\
     & & \\
     & & \\
     & &
\end{quantikz}
=
\begin{quantikz}[row sep = 0.15]
     & \gate[2]{G_1} & & \gate[2]{G_3}& & \\
     & & \gate[2]{G_2} & & \gate[2]{G_4} & \\
     & \gate[2]{G_1} & & \gate[2]{G_3}& & \\
     & & \gate[2]{G_2} & & \gate[2]{G_4} & \\
     & \gate[2]{G_1} & & \gate[2]{G_3}& & \\
     & & & & & 
\end{quantikz}
\end{equation*}
 
RTR is applied to the manifold $\mathcal{M}=U(4)^m$ with the Riemannian metric induced by its standard embedding in $(\mathbb{C}^{4\times 4})^m$ together with a retraction defined by the QR-decomposition. In the case $m=1$, given $(G,H)\in T\mathcal{M}$, we take $R(G,H)=qf(G+H)$, where $qf(A)$ for $A\in\mathbb{C}^{4\times 4}$ denotes the unitary matrix in the QR-decomposition $A=qf(A)\cdot R$ with $R\in\mathbb{C}^{4\times 4}$ upper triangular with strictly positive diagonal entries. This definition then extends in the natural way to the cases $m>1$. Finally, we take the cost function to be $f(G)=\|E(G)-U\|_F^2$, where $F$ is the Frobenius norm distance in $U(2^n)$, a measurement of the error of our brick-wall approximation $E(G)$ away from $U$, which we seek to minimise. 

\begin{figure}[H] 
    \centering
    \includegraphics[height=54mm]{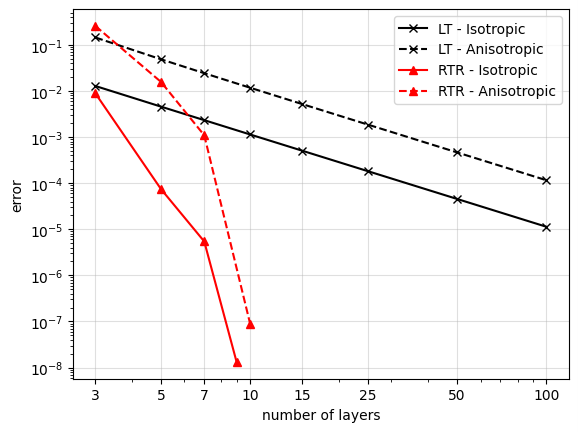}
    \caption{Plot of error as a function of the number of layers of the Hamiltonian circuit for Lie-Trotter-Suzuki (LT) and Riemannian Trust Regions (RTR), for two different parameter values on 5 qubits systems at $t=1$.\label{RTR_LT}}
\end{figure}

RTR can be more generally used to optimise over a Stiefel manifold to map Hamiltonian systems onto circuits~\cite{HVMH21} and using standard libraries~\cite{manopt}. Previous studies have advocated the usefulness of RTR method over product formulas~\cite{KBHM24}, and exemplified in \Cref{RTR_LT}.

\section{Results}\label{Results}
Having described the YKY-RTR algorithm in the previous sections, \ref{SEC:YKY}, \ref{SEC:RTR}, we now report the performance of this algorithm to compute OTOCs and effectively measure the butterfly velocity of the XY model on current NISQ devices. 

In particular, we implement the YKY-RTR algorithm on IBM-Q quantum devices, and use the noisy quantum simulator (FakeTorino) to obtain our results. We use the quantum simulator to first measure OTOCs on a 5-qubit system \eqref{H_XY}, for parameters in \Cref{Table:Params}, and compute squared commutator $C_j(t)$ by first projecting qubits $A_j, B_j$ to $\dmat{\phi}_{A_j B_j}$ and then measuring qubits $A_0, B_0$ in the Bell basis and recording the probability of observing the outcomes $"00"$ as $F_{EPR}$. Subsequently we estimate $C_j(t) = 2-\frac{1}{2F_{EPR}}$.

\begin{widetext}

\begin{figure}[H]
    \centering
        \subfloat[Isotropic Wireframe]{\includegraphics[width=0.3\textwidth]{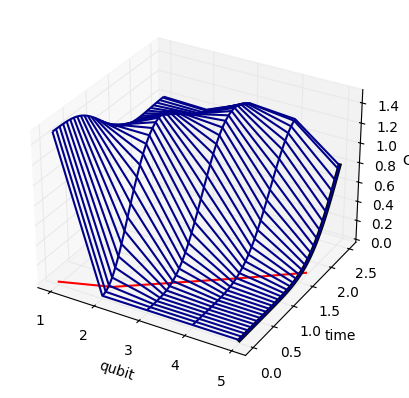}\label{iw}}
        \subfloat[Isotropic $C_5(t)$]{\includegraphics[width=0.3\textwidth]{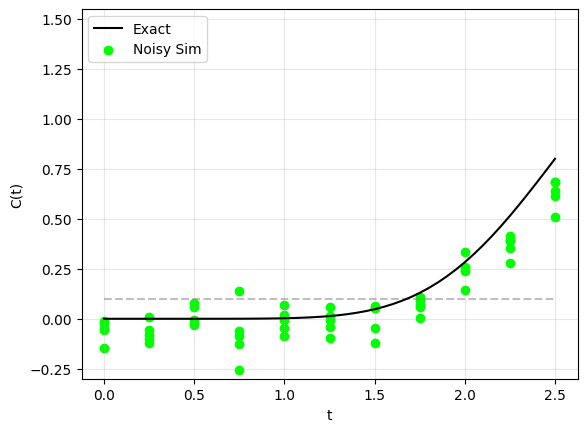}\label{ic}}
        \subfloat[Isotropic Butterfly Velocity]{\includegraphics[width=0.3\textwidth]{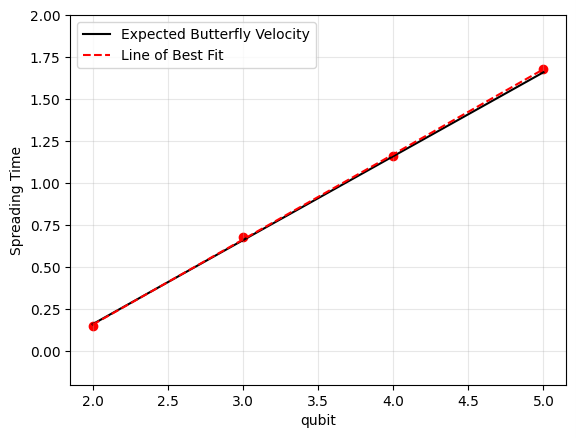}\label{ibv}}\\
        \subfloat[Anisotropic Wireframe]{\includegraphics[width=0.3\textwidth]{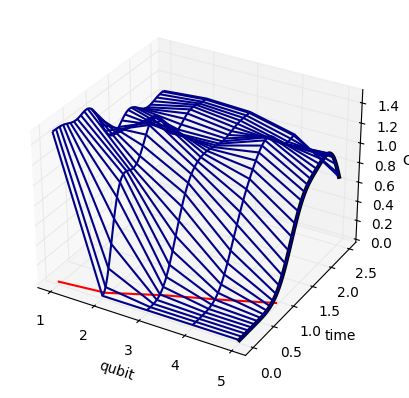}\label{aw}}
        \subfloat[Anisotropic $C_5(t)$]{\includegraphics[width=0.3\textwidth]{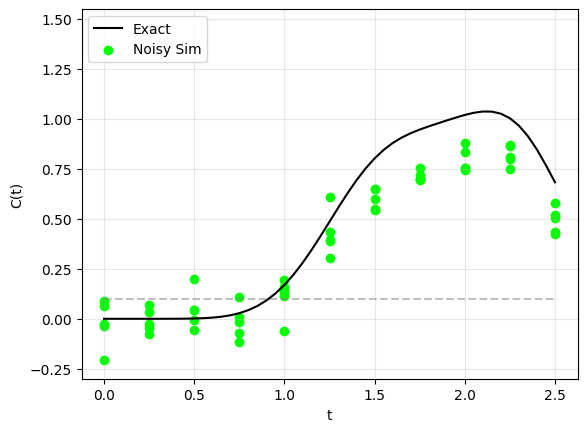}\label{ac}}
        \subfloat[Anisotropic Butterfly Velocity]{\includegraphics[width=0.3\textwidth]{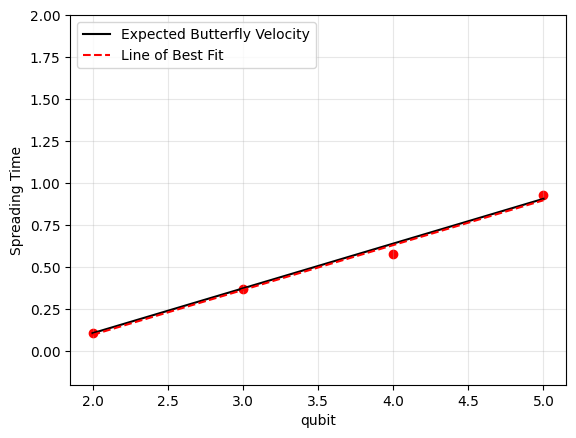}\label{abv}}
    \caption{Plots demonstrating proof-of-concept for the YKY-RTR algorithm on parameter sets in \Cref{Table:Params}. Figures (a) and (d) demonstrate noiseless wireframe plots of the squared commutator $C_j(t)$ as a function of qubit position $j$ and time $t$. The red line in the $C_j(t) = 0$ plane shows the spreading time $t_j$ as a function of $j$. Figures (b) and (e) show noiseless plots of the commutator $C_5(t)$ as black lines with dots representing noisy quantum simulations. The dashed grey line represents the bound for spreading time $C_5(t) = 0.1$. Figures (c) and (f) show the process of finding the butterfly velocity from the spreading times (calculated again via noisy quantum simulation).}
\end{figure}

\end{widetext}

We first observe the idealised (noiseless) values $C_j(t)$ as a function of $j$ and $t$ on the 5 qubit Hamiltonian. This data is presented using the wireframe in \Cref{iw} and \Cref{aw}, with qubit position and time on the $x$ and $y$ axes respectively, and the $z$ axis displaying $C_j(t)$. This then gives a quantitative presentation of the operator spreading in the XY model. The red line in the $C_j(t)=0$ plane marks the spreading time $t_j$ for each qubit position $j$, which is defined as $t_j = \displaystyle\min_t \{C_j(t) \geq 0.1\}$.

To demonstrate the efficacy of YKY-RTR in obtaining such results on current NISQ-era quantum computers, we then simulate the YKY-RTR algorithm on a noisy quantum computer and estimate the value of $C_5(t)$ for the given parameters. In \Cref{ic} and \Cref{ac} we report our findings of the values of $C_5(t)$ obtained from the noisy simulation on IBM-Q FakeTorino simulator, that is otherwise expected to predict the outcomes of the real device. 

We then estimate $C_j(t)$ for the various $j \in \{2,3,4,5\}$, record the data and obtain the spreading time $t_j$ as in the classical wireframe plots. A line of best fit is then calculated using least squares method, and the reciprocal of the slope of this line then gives a numerical approximation for the butterfly velocity. Plots for these are included in \Cref{ibv} and \Cref{abv}. These approximations for the butterfly velocity (calculated using both classical and quantum simulations) are shown in \Cref{Table:Params}.

\begin{table}[ht]
\centering
\begin{tabular}{|c|c|c|c|c|c|c|}    \hline
                      & J                     & r   & h   & True vB & Numerical & Quantum  \\ 
                      &                       &     &     &         & 5 qubits      & 5 qubits \\     \hline
 %\cline{2-2}
Isotropic             & 1                     & 0   & 0   & 2       &  2.066             & 1.972    \\        \hline
Anisotropic                 & 1                     & 2.1   & 0.8 &  3.75      &   3.717         &   3.745  \\        \hline
\end{tabular}
    \caption{Parameter choices for XY Model}
    \label{Table:Params}

\end{table}

\section{Discussions}
Here we investigated the efficacy of the YKY-RTR method to measure growth of commutators on current generation quantum computers, and subsequently measure the butterfly velocity in spin systems. We then compare the numerical predictions obtained using the YKY-RTR method on IBM-Q noisy devices (FakeTorino) against the analytical solutions and found them to be in agreement on instances as small as $n=5$ qubit systems. For generic lattice systems, larger $n$ would be sufficient. 

Our findings demonstrate the effectiveness of YKY algorithm to robustly estimate OTOC and highlights the usefulness of the manifold optimisation of Riemannian-Trust-Region Method for Hamiltonian to circuit mapping that are significantly more efficient than product formulas, in the non-asymptotic sense. Our investigation complements previous studies on measuring OTOCs, such as in \cite{SBSH16, ZHG16, Yao16, Google21,Zoe_2022} and references therein.

The focus of this study was to understand and report the efficacy of ab-initio scalable methods for quantum computations of butterfly velocity. We do not use error mitigation techniques but rely on internal robustness of the algorithm. The YKY-RTR method can be more generally useful to calculate butterfly velocity and speed of information propagation in systems that may not be analytically solvable. 
\subsection*{Acknowledgements}
\noindent CM and SRM thank Cambridge Summer Research in Mathematics program. EC was partially supported by NSF Award DMS-3103392 as well as SLMath (MSRI) during Fall 2024 (partially supported by the NSF Grant DMS-1928930). SRM is supported by “Quantum
simulation algorithms for quantum chromodynamics” grant (ST/W006251/1).

\bibliographystyle{unsrt}
\bibliography{bibliography}

\newpage
\appendix
\section{Analytic Calculation for Butterfly Velocity}
\label{SEC:APPENDIX}
Here we will provide a more detailed explanation of the calculations done in Section \ref{SEC:Analytic}. We start with the spin lattice Hamiltonian, included here for convenience.

\begin{equation}
    H = J\sum_j \left(\frac{1+r}{2}X_jX_{j+1} +\frac{1-r}{2} Y_jY_{j+1} + hZ_j \right)
\end{equation}

The first step is to convert this spin lattice Hamiltonian into a fermionic Hamiltonian via a Jordan-Wigner transform. Here we use the following convention to perform this transformation:

\begin{align}
    X_j &= - \prod_{k<j}(\mathbb{I}-2f_k^{\dagger} f_k)(f_j+f_j^{\dagger}) \nonumber \\
    Y_j &= -i \prod_{k<j}(\mathbb{I}-2f_k^{\dagger} f_k)(f_j-f_j^{\dagger}) \\
    Z_j &= \mathbb{I}-2f_j^{\dagger} f_j \nonumber
\end{align}

For completeness, $f_j$ can be explicitly derived from (A2) by substituting the equation for $Z_k$ into the equations for $X_j$ and $Y_j$ and using the fact that Pauli matrices are hermitian and unitary (i.e. $Z_k^2 = \mathbb{I}$). Solving the equations simultaneously gives $f_j = Z_1Z_2...Z_{j-1}a_j$ with $a_j = \mathbb{I}_{2^{j-1}} \otimes \begin{pmatrix} 0 & 1 \\ 0 & 0 \end{pmatrix} \otimes \mathbb{I}_{2^{n-j}}$ the standard annihilation operator on site $j$.

The non-local terms in $X_j$ and $Y_j$ are a phase term dependent on the number of fermions on sites prior to $j$, giving a $-1$ phase if this number is odd and a $+1$ phase if it is even.

To satisfy the conditions for a fermionic Hamiltonian, these fermionic annihilation and creation operators $f_j$ and $f_j^{\dagger}$ must satisfy certain anti-commutation relations. In particular, $\{f_i, f_j^{\dagger} \} = \delta_{i, j} \mathbb{I}$ and $\{f_i, f_j \} = 0$. This second relation also gives us that $(f_j)^2 = 0$, which demonstrates the Pauli-exclusion principle, as trying to destroy 2 qubits on the same site means destroying the whole system.

Plugging this into equation (A1) requires the calculation of $X_jX_{j+1}$ and $Y_jY_{j+1}$ in terms of these new operators. The calculation will be done here for $X_jX_{j+1}$, but follows similarly for the $Y_jY_{j+1}$ case. Writing the equation out in full gives:

\begin{align}
    X_jX_{j+1} &= \left(\prod_{k<j} (\mathbb{I}-2f_k^{\dagger}f_k)\right)(f_j+f_j^{\dagger}) \cdot\\
    & \quad \left(\prod_{k<j+1} (\mathbb{I} -2f_k^{\dagger}f_k)\right) (f_{j+1} + f_{j+1}^{\dagger}) \nonumber
\end{align}

From this second product we can pull out the $j$ term, and then use the anti-commutation laws to move terms around. The $(f_j+f_j^{\dagger})$ term will anti-commute with $f_k$ and $f_k^{\dagger}$ for $k<j$. Hence by anti-commuting it with both, it must commute with $f_k^{\dagger}f_k$. Thus we can swap these terms. By this same logic, we can swap terms within each product with each-other, so the equation becomes:
\begin{equation}
    \hspace{-0.4cm} X_jX_{j+1} = \left(\prod_{k<j} (\mathbb{I}-2f_k^{\dagger}f_k)^2\right) (f_j+f_j^{\dagger})(\mathbb{I}-2f_j^{\dagger}f_j)(f_{j+1} + f_{j+1}^{\dagger})
\end{equation}

From a physical perspective, the product term is expected to cancel due to double-counting. Since it provides only a phase, which depends on the parity of the number of fermions, and each fermion in the product is counted twice, there is overall an even number of fermions here, so the phase factor is just $+1$. Algebraically this can also be seen via:

\begin{align}
    (\mathbb{I}-2f_k^{\dagger}f_k)^2 &= \mathbb{I}-4f_k^{\dagger}f_k+4f_k^{\dagger}f_kf_k^{\dagger}f_k \nonumber \\ 
    &= \mathbb{I}-4f_k^{\dagger}f_k + 4f_k^{\dagger}f_k - 4(f_k^{\dagger})^2(f_k)^2\\
    &= \mathbb{I} \nonumber
\end{align}

where between the first and second lines we have used the anti-commutation relation $f_kf_k^{\dagger} = \mathbb{I}-f_k^{\dagger}f_k$, and between the second and third lines we have used the fact that $(f_k^{\dagger})^2 = (f_k)^2 = 0$. A similar process can be applied to the $j$ terms in (A4), giving the final equation for $X_jX_{j+1}$ as:
\begin{equation}
    X_jX_{j+1} = (f_j^{\dagger}-f_j)(f_{j+1} + f_{j+1}^{\dagger})
\end{equation}

Performing a similar process for $Y_jY_{j+1}$ then subbing both expressions into (A1) (along with the $Z_j$ term directly from (A2)) gives the fermionic Hamiltonian:
\begin{align}
    H = J\sum_{j} &r(f_{j+1}f_{j} + f_j^{\dagger}f_{j+1}^{\dagger}) + (f_{j+1}^{\dagger} f_j + f_j^{\dagger} f_{j+1}) \nonumber \\
    &+ h (\mathbb{I}-2f_j^{\dagger} f_j) 
\end{align}

The next step is to perform a Fourier Transform from position space into the momentum space, which converts our localised position operators $f_j$ into non-local momentum operators $c_k$ via $f_j  =  \sum_{k=1}^n c_k e^{ikj}/\sqrt{n}$. We can substitute this in for each term in (A7) then move the summations around to cancel terms down. This can be done relatively easily for terms conserving fermion number such as $f_j^{\dagger}f_{j+1}$ as:

\begin{align}
    \sum_j f_j^{\dagger}f_{j+1} &= \frac{1}{n} \sum_{j, k, l}c_k^{\dagger}c_{l}e^{ikj}e^{-il(j+1)} \nonumber \\
    &= \frac{1}{n} \sum_{k, l} c_k^{\dagger}c_l e^{-il} \underbrace{\sum_j e^{ij(k-l)}}_{n\delta_{k l}}\\
    &= \sum_k c_k^{\dagger}c_k e^{-ik} \nonumber
\end{align}

The summation in the second line occurs due to roots of unity if $k \neq l$, and simplifies the expression considerably. A similar idea can be done for the terms multiplied by $r$ in (A7), with one extra step to leave a trig function instead of complex exponentials.
\begin{align}
    \sum_j f_{j+1}f_j &= \frac{1}{n} \sum_{j, k, l}c_kc_le^{ik(j+1)}e^{ilj} \nonumber \\
    &= \frac{1}{n}\sum_{k, l} c_kc_l e^{ik} \underbrace{\sum_j e^{ij(k+l)}}_{n\delta_{-k l}} \\
    &= \frac{1}{2}\sum_k c_kc_{-k}e^{ik}+ \frac{1}{2}\sum_l c_{-l}c_l e^{-il} \nonumber \\
    &= \frac{1}{2}\sum_k c_kc_{-k}(e^{ik}-e^{-ik}) =  i\sum_k c_kc_{-k} \sin{k} \nonumber
\end{align}

Between the second and third line, we have used the fact that we can either replace $l$ with $-k$ or replace $k$ with $-l$, to split the summation in half. Then, since $k$ and $l$ are just dummy variables, we can replace the $l$ with $k$ in the third line. Crucially, $c_k$ and $c_{-k}$ anti-commute, so we can swap these terms in the second sum on the third line and pick up a minus sign to reach the fourth line.

The processes in (A8) and (A9) can be repeated for the remaining terms in (A7) to get:
\begin{align}
     H = -J \sum_k & \enspace 2(h-\cos(k)) c_k^{\dagger}c_k \\
     &+ ir\sin(k) (c_{-k}^{\dagger}c_k^{\dagger} + c_{-k}c_k)-h\mathbb{I} \nonumber
\end{align}

The final step required is to perform a Bogoliubov transform, to diagonalise (A10). This requires a new fermionic operator $\gamma_k = u_k c_k -iv_kc_{-k}^{\dagger}$ with $u_k, v_k \in \mathbb{R}$. To ensure this operator satisfies the fermionic anti-commutation relations requires $u_k^2+v_k^2=1$, $u_{-k} = u_k$ and $v_{-k} = -v_k$. Because of this, we choose to write $u_k = \cos(\theta_k/2)$ and $v_k = \sin(\theta_k/2)$ with $\theta_k$ the Bogoliubov angle.

To transform (A10) using this new operator we need the inverse transform, which is written $c_k = u_k \gamma_k + iv_k\gamma_{-k}^{\dagger}$. We substitute this equation into (A10), and use the anti-commutation relations and double angle formulae with $\theta_k$, to get:
\begin{align}
    H  = &-J \sum_k \gamma_k^{\dagger}\gamma_k[2(h-\cos{ka})\cos^2{\frac{\theta_k}{2}}+r\sin{ka}\sin{\theta_k}] \nonumber \\
    &+ \gamma_{-k}\gamma_{-k}^{\dagger}[2(h-\cos{ka})\sin^2{\frac{\theta_k}{2}}-r\sin{ka}\sin{\theta_k}] \\ \nonumber
    &+ i(\gamma_{-k}\gamma_{k} + \gamma_{-k}^{\dagger}\gamma_k^{\dagger})(r\sin{ka}\cos{\theta_k} -(h-\cos{ka})\sin{\theta_k}\\ 
    &-h\mathbb{I} \nonumber
\end{align}

To diagonalise (A11), we need to remove the third term, which can be done by taking $\tan{\theta_k} = r\sin{ka}/(h-\cos{ka})$. We then note that $\sum_k = \sum_{-k}$ due to periodicity, so we can swap $-k$ to $k$ in the second term. Then, we can use the anti-commutation relation $\gamma_k\gamma_k^{\dagger} = \mathbb{I} - \gamma_k^{\dagger}\gamma_k$, to reach:
\begin{align}
    H  = -J \sum_k &\gamma_k^{\dagger}\gamma_k \cdot 2[(h-\cos{ka})\cos{\theta_k}+r\sin{ka}\sin{\theta_k}] \nonumber \\
    &-h + 2(h-\cos{k})\sin^2{\frac{\theta_k}{2}} -r\sin{\theta_k}\sin{k}
\end{align}

Plugging in the value of $\theta_k$ (and noting that $\sum_k \cos{k} = 0$ due to periodicity) we reach the final expression:
\begin{equation}
    H = \sum_k \varepsilon(k;r,h)\left(\gamma_k^{\dagger}\gamma_k - \frac{1}{2}\right)
\end{equation}

with the energy dispersion relation:
\begin{equation}
    \varepsilon(k;r,h) = -2J\sqrt{(h-\cos{k})^2+r^2\sin^2{k}}
\end{equation}

\end{document}